\documentclass[]{interact}

\usepackage{epstopdf}
\usepackage[caption=false]{subfig}
\usepackage{graphicx}
\usepackage{multirow}
\usepackage{url}
\usepackage{tikz}
\usepackage{lmodern}

\usepackage[longnamesfirst,sort]{natbib}
\bibpunct[, ]{(}{)}{;}{a}{,}{,}

\theoremstyle{plain}

\theoremstyle{definition}

\theoremstyle{remark}

\newcommand{\UMFPACK}{\textsc{Umfpack}}
\newcommand{\Triangle}{\textsl{Triangle}}
\newcommand{\UofA}{{University~of~Arkansas}}
\newcommand{\anon}[1]{{}}
\begin{document}


\title{Training the next generation of computational scientists through a new undergraduate course}

\author{
\name{Tulin Kaman\textsuperscript{a}\thanks{CONTACT Tulin Kaman. Email: tkaman@uark.edu}, 
Rouben Rostamian\textsuperscript{b} 
and Shannon W. Dingman\textsuperscript{a}}
\affil{\textsuperscript{a} Department of Mathematical Sciences, University of Arkansas, Fayetteville, AR 72701, USA; \textsuperscript{b}Department of Mathematics and Statistics, University of Maryland, Baltimore County, Baltimore, MD 21250, USA}
}

\maketitle

\begin{abstract}

We introduce a newly designed undergraduate-level
interdisciplinary course in scientific computing that
aims to prepare students as the next generation of
research-oriented computational scientists and engineers.
The course offers students 
opportunities to explore a diverse set of projects and develop 
the necessary programming skills to implement ideas and algorithms 
within high performance computing environments. 
The training includes how to think about, formulate, 
organize, and implement programs in scientific computing. 
The emphasis of the course is on problem solving within a wide range of 
applications in science and engineering.  

\end{abstract}


\section{Introduction}
\label{sect:introduction}

Computational Science and Engineering (CSE) is a rapidly growing field. 
Theory and experiments are supported by computational work to understand 
the behavior of complex systems arising in science and engineering applications. 
The set of knowledge and skills needed in CSE lies at the intersection of 
mathematics, computer science and natural sciences and engineering.  
Many scientific and engineering problems are described by mathematical models, 
the models are analyzed and solved using numerical algorithms, which in turn 
are implemented using programming languages. The development of efficient, 
accurate and robust software for the numerical simulation of complex systems 
is the key point in CSE research.  The current trends on CSE is
are captured through the following indicators:

\begin{itemize}
\item
The U.S. Department of Energy (DOE) Advanced Scientific Computing 
Research (ASCR) (\url{https://science.energy.gov/ascr}) program states the importance of CSE:  
``It is generally accepted that computer modeling and simulation offer 
substantial opportunities for scientific breakthroughs that cannot 
otherwise---using laboratory experiments, observations, or traditional 
theoretical investigations---be realized. At many of the research frontiers, 
computational approaches are essential to continued progress and play 
an integral and essential role in much of twenty-first century science and engineering."

\item
Officers of the Society for Industrial and Applied Mathematics (SIAM) 
Activity Group on Computational Science and Engineering~\citep{RudeWillcox18} 
examine the role of CSE in the 21st-century and 
discuss the challenges and opportunities in CSE research including 
mathematical methods and algorithms, high performance computing, and data science.  

\item
The ASCR Scientific Discovery through Advanced Computing (SciDAC)
\url{https://scidac.gov/}
program brings computational scientists, applied mathematicians, and computer scientists 
together with the mission to develop the scientific computing software and hardware 
infrastructure needed to advance scientific discovery using supercomputers. 
Two SciDAC institutes, (i) FASTMath -- Frameworks, Algorithms, and Scalable 
Technologies for Mathematics (\url{https://scidac5-fastmath.lbl.gov}) 
and (ii) RAPIDS -- SciDAC Institute for Computer Science and Data (\url{https://rapids.lbl.gov/}) 
are designed to develop scientific advances in petascale computing. 
The ASCR Applied Mathematics program supports research on numerical methods, 
optimization, multiphysics-multiscale computation and math software development 
for the numerical studies in computational fluid dynamics, climate modeling, nuclear 
reactor design, subsurface flow modeling, and many other applications. 
\end{itemize}

While advanced computational efforts are growing rapidly, it is important to adapt 
student education to keep up with the requirements of a career in academia and industry. 
Therefore, there have been many undergraduate student opportunities provided 
at universities and national laboratories/facilities supported by National Science 
Foundation (NSF) Research Experience for Undergraduates 
(REU) programs (\url{https://www.nsf.gov/crssprgm/reu}) and 
U.S. Department of Energy's Science Undergraduate Laboratory 
Internships (SULI) (\url{https://science.osti.gov/wdts/suli}). 
Moreover professional organizations such as the American Mathematical Society (AMS), 
the Association for Women in Mathematics (AWM), and Society for Industrial 
and Applied Mathematics (SIAM) provide support for undergraduate students' developments. 
The SIAM working group on CSE undergraduate education~\citep{ShifletVakalis11} 
stresses the importance of CSE, the need of training in CSE fundamentals at 
undergraduate level and the skills needed for the training of students. 

We believe that scientific computing should be an essential part of students' 
education if they are to be active  participants in the development of future technology.
Specifically, it is important for students to be able to develop an understanding 
of the methods, algorithms, and computing skills that are necessary for scientific computing
during their undergraduate studies. Therefore, we proposed a new course, 
MATH~4343 ``Introduction to Scientific Computing" in the Department of 
Mathematical Sciences at the \UofA\ which was taught 
in Spring 2021 for the first time.
The learning objectives of the MATH~4343 are divided into three parts:
{
\renewcommand{\theenumi}{\roman{enumi}}
\begin{enumerate}
  \item understanding the problem with its mathematical model;
  \item learning the algorithms needed for solving the mathematical models; and 
  \item the implementation of algorithms in the Linux environment.
\end{enumerate}
}
 MATH~4343 provides students with a theoretical and practical background 
 sought after in the industry by introducing a diverse set of real world 
 problems in science and engineering such as the Nelder--Mead 
 downhill simplex, the heat equation, the porous medium equation, 
 and the Finite Elements Method~\citep{Rostamian14}. 
 The emphasis of MATH~4343 is on problem solving, and offers 
 multiple projects based on the students'
 backgrounds. This course was originally designed for the students 
 of Applied Mathematics in the Department of Mathematical Sciences 
 to help them quickly acquire mathematical and programming skills 
 to solve a diverse set of problems from a variety of engineering areas. 
 Meanwhile, it has gained popularity among other science and engineering 
 students at the \UofA. 
 
 MATH~4343's main goal is to motivate and prepare undergraduate students 
 for their role as future computational scientists and engineers. The course 
 is designed to support programs with computational concentration, 
which recommend or require certain level of programming expertise to students. 
To take advantage of the current computational resources provided at the university, 
the students become familiar with C programming language in the Linux environment 
and auxiliary software packages and libraries needed for visualizing. The prerequisite 
for this course is only linear algebra which is the highest level mathematics course 
taken by most of the natural sciences and engineering students; however no 
previous programming experience in C is required.
The time required to teach C programming from basics to a moderately advanced 
level is built into the course design, and meshes well with the additional activities 
through lecture series on programming (see section~\ref{sect:activities}).
 
In section~\ref{sect:details} we describe the course's details, including the content 
(section~\ref{sect:content}) and performance assessments (section~\ref{sect:perfAsses})
which are designed based on the enrolled students' needs and backgrounds. 
In section~\ref{sect:activities} we present a list of activities designed to achieve
 the course's main goal and address the importance of additional learning 
 activities in the form of seminars and workshops to train and prepare 
 students for their future careers.  In section~\ref{sect:internships} 
 we present the opportunities specifically presented to 
 students who complete the course and wish to conduct research.  
 Student evaluation of the course and instructor are presented in section ~\ref{sect:eval}.
 Finally, we present the concluding remarks and future plans in section~\ref{sect:conclusion}.
 
\section{Course Details}
\label{sect:details}
In this section, we present the three primary components of the course, which
consist of
i)~\emph{the learning objectives},
ii)~\emph{assessments}, and
iii)~\emph{instructional strategies}.
The learning objectives component is designed with the purpose of 
guiding the student on how to interpret and analyze
a diverse set of problems derived from science and engineering,
and then devise and implement algorithms to produce quantitative
information about them.
The skills learned here provide a foundation
upon which other courses within the scientific computing domain can build. 

We examine the relative efficiency of various algorithms
applied to a given problem, such as explicit, implicit, hybrid
finite difference schemes~\citep{isaacson-keller} for solving initial/boundary problems
associated with the classical heat equation.
The understanding gained here paves the way to handling more complex problems
such as the porous medium equation~\citep{vazques-pme-book} which arises in
the study of diffusion of gasses, and also in population
dynamics~\citep{gurtin-maccamy}.

In our experience, a solid knowledge of undergraduate multivariable calculus and
linear algebra provides sufficient mathematical background for this course.
We have had success with setting MATH~3093 (Abstract Linear Algebra)
for mathematics students, or MATH~3083 (Linear Algebra) for engineering
students as the prerequisite for this course.

Most of the undergraduate students who enroll in this course
have had prior exposure to elements of programming through
Python and/or MATLAB, or perhaps lower level languages such as
C and Java.  Students who lack experience with C or a C-like
procedural programming language, can still succeed through
self-study and some extra help from the instructor and their
classmates.

On the first day of the class we administer a survey to learn about the
students' backgrounds.
The survey questions are based on two of the
\textsl{Mathematics Attitudes and Perceptions Survey} (MAPS)~\citep{MAPS} categories:
\emph{interest} and \emph{expertise}.
The feedback from the survey helps the instructor to design the
project-based course content including a two-week\footnote{
	C is a particularly compact programming language and the bulk of it
	may be learned rather quickly.
}
crash course on the C programming language,
and the programming environment on the university's high performance
computing system, Pinnacle.
It's safe to say that upon completion of this course,
the students will have been exposed to about
90\% of the C programming language features.

\begin{figure}
\centering
\begin{tikzpicture}[{every node/.style={black,font=\sffamily\large}}, scale=0.9]
    \def\firstcircle{(0,0) circle (3cm)}
    \def\secondcircle{(3,0) circle (3cm)}
    \def\thirdcircle{(1.5,3) circle (3cm)}
    \def\boundingbox{(-3,-3) rectangle (6,4.5)}

    \definecolor{science}{HTML}{c4c7da}
    \definecolor{sciengr}{HTML}{d7c4da}
    \definecolor{sciengrmath}{HTML}{e4d5b3}
    \definecolor{scimath}{HTML}{c4dad7}
    \definecolor{engr}{HTML}{c7dac4}
    \definecolor{engmath}{HTML}{dad7c4}
    \definecolor{math}{HTML}{dac4c7}
 
    \fill[engr] \firstcircle node[align=center,xshift=-1.5cm, yshift=-.5cm] {Engineering \\ 2};
    \fill[math] \secondcircle node[align=center,xshift=1.5cm, yshift=-.5cm] {Mathematics \\ 1};
    \fill[science] \thirdcircle node[align=center,yshift=1cm] {Science \\ 0};

    \begin{scope}
        \clip \boundingbox \thirdcircle;
        \clip \firstcircle;
        \fill[engmath] \secondcircle node[black, xshift=-1.5cm, yshift=-1cm] {2};
    \end{scope}
    \begin{scope}
        \clip \boundingbox \secondcircle;
        \clip \firstcircle;
        \fill[sciengr] \thirdcircle node[xshift=-1.8cm, yshift=-1cm] {2};
    \end{scope}
    \begin{scope}
        \clip \boundingbox \firstcircle;
        \clip \secondcircle;
        \fill[scimath] \thirdcircle node[xshift=1.8cm, yshift=-1cm] {3};
    \end{scope}
    \begin{scope}
        \clip \firstcircle;
        \clip \secondcircle;
        \clip \thirdcircle;
        \fill[sciengrmath] \boundingbox;
    \end{scope}

    \node at (1.5,1.2) {0};
\end{tikzpicture}
\caption{Number of students for each one or combination of the disciplines from 
science, engineering and mathematics.}
\label{fig:studentsBack}
\end{figure}
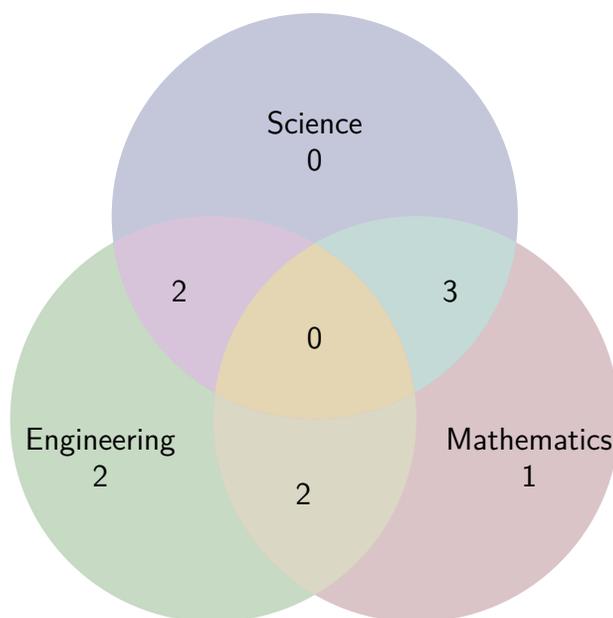

There were ten students enrolled in Spring 2021 when
MATH~4343 was offered for the first time. 
Twenty-five percent of the students who enroll in this course
are juniors who have earned between 60 and 90 credit hours. 
The rest are seniors who have earned 90 or more credit hours.
Figure~\ref{fig:studentsBack} displays these ten students' major areas of study.
On the first day of class, a survey was given to learn more 
about the students' interests and learning preferences. 
See a summary of survey responses for each question in Table~\ref{interests}. 

\begin{table}[htbp]
\tbl{Student interests and learning survey summary}
{\begin{tabular}{ |p{5cm}|l|c| } 
  \hline
 Questions & Answers & Responses \\  
  & & (\%) \\  \hline 
 \multirow{3}{*}{\parbox{4.75cm}{When I am learning something new, I prefer to}}
 & have someone \textbf{\textsl{show}} me &  70\%	  \\ 
 & have someone \textbf{\textsl{tell}} me & 20\%  \\  
 & figure it out myself & 10\% \\  \hline
  \multirow{2}{*}{\parbox{4.75cm}{I prefer to work }}
 & individually &  50\%	  \\ 
 & with a group & 50\%  \\  \hline
 \multirow{3}{*}{\parbox{4.75cm}{What is your level of programming? }}
 & beginner &  30\%	  \\ 
 & intermediate  & 70\%  \\ 
 & advanced & 0\%  \\  \hline
 \multirow{5}{*}{\parbox{4.75cm}{How experienced are you in C programming? }}
 &novice &  50\%	  \\ 
 & advanced beginner & 40\%\\
 & competent & 10\% \\
 & proficient & 0\%  \\ 
 & expert & 0\% \\ \hline 
  \multirow{5}{*}{\parbox{4.75cm}{How experienced are you in Linux environment? }}
 & novice &  40\%	  \\ 
 & advanced beginner & 40\%\\
 & competent & 0\% \\
 & proficient & 20\%  \\ 
 & expert & 0\% \\ \hline
  \multirow{5}{*}{\parbox{5cm}{How many programming based courses have you
  taken? }}
 & 1 & 20\% \\ 
 & 2 & 40\% \\ 
 & 3 & 10\% \\ 
 & 4 & 20\% \\ 
 & 5+more & 10\% \\ \hline
 \multirow{2}{*}{\parbox{4.75cm}{Have you been involved in undergraduate research? }}
 & yes &  30\%	  \\ 
 & no & 70\%\\ \hline
 \multirow{2}{*}{\parbox{4.75cm}{If not,  are you interested in acquiring research experience?  }}
 & yes &  0\%	  \\ 
 & no & 100\%\\ \hline
\end{tabular}}
\label{interests}
\end{table}

\subsection{Course content}
\label{sect:content}

The course content and its pace are adjusted 
to what typical undergraduate students can handle.
In the following subsections we list the typical course content.

\subsubsection{A two-week crash course on programming and computing environments}

The B.S. degree in Natural Sciences with Computational Concentration programs
requires core courses that include Calculus I--II, Elementary Differential
Equations, Linear Algebra, as well as computer science courses such as
Programming Foundations I--II and Programming Paradigms.

Some other engineering and science programs recommend, but do not
require, courses in programming as degree requirements.
To reach as wide an audience
as possible, and to establish a common background for all
enrolled students in MATH~4343 course, we begin with the basics
concepts of C programming (data representation, conditional
and iterative statements, functions, arrays, strings), and then
continue with more advanced topics of C (pointers, structures,
dynamic allocations, recursion, linked lists, binary trees,
unions and function pointers).
The bare minimum of the C programming language introduced
this way is mostly adequate for the average student.  To the
more ambitious students, who desire further reading and a comprehensive
reference, we recommend the book by Kochan~\citep{kochan-2014} which 
integrates well with this course's objectives.

We introduce the \textsl{Arkansas High Performance Computing
Center} (AHPCC) computing environment which is used throughout
this course.  AHPCC provides high performance computing
hardware, storage, support services, and training to enable
computationally-intensive and data-intensive research.
The AHPCC is available to faculty, staff and students at all of
the Arkansas public universities, and supports educational
allocations which are used in teaching this course.
The undergraduate students first get familiar with the Linux
system and the command-line based terminal.
This was a new experience for all but one of the students 
who had used the system previously in another research project.

A benefit of using the AHPCC is the creation of a uniform computing
environment for all students, as they don't need to download
and install compilers and third-party software packages in
order to work on their programming projects.
The students can easily switch between different compilers 
such as the GNU and Intel C compilers and experiment
with different software tools for debugging, performance analysis, etc. 
In addition, third-party software packages such
as \textit{Geomview}, \Triangle, and \UMFPACK\ are
readily available for their use.

After the two-week crash course on programming and computing environments, 
we closely follow the textbook
``\textit{Programming Projects in C for Students of Engineering, Science and 
Mathematics}"~\citep{Rostamian14} throughout the semester. 
This book is designed for early graduate students and advanced undergraduate students. 
The textbook consists of two parts. The six chapters of
Part~I (A Common Background) are prerequisites for Part~II (Projects) which we now describe.

\subsubsection{Projects}
\label{sec:proj}

The identification and formulation of engineering problems via 
mathematical modeling, the solution of problems using numerical methods, 
and the design of computer programs, are the fundamental blocks of 
training for students who plan to pursue a career as computational 
scientists and engineers. We therefore begin with identifying a list of projects 
that are relevant to the students' backgrounds. 

The projects presented in the textbook are interesting, intriguing, inviting, challenging,
and illuminating on their own, apart from their programming aspects.
Some of the favorite projects in the textbook are:
\begin{enumerate}
\item The {\it Nelder--Mead simplex algorithm}~\citep{1965-nelder-mead,NRC} for minimizing functions, with application to computing finite deformations of trusses under large loads via minimizing the energy, as well as training neural networks;
\item The {\it Haar wavelet transform}, with application to image analysis and
image compression~\citep{wavelet-book-nievergelt,stollnitz-etal-1995a,wavelet-book-stollnitz-etal};
\item {\it The evolution of species} and the effect of the environment on the
emergence of genetically distinct species~\citep{Dewdney1989,LandOfLisp};
\item {\it Finite difference (FD) algorithms} for solving the time-dependent linear
heat equation and extending the algorithms to solve the porous medium
equation~\citep{isaacson-keller,Rostamian14};
\item {\it Finite element methods (FEM)} for solving second order elliptic partial differential equations on arbitrary two-dimensional domains through
unstructured triangular meshes and linear elements~\citep{fem-szabo-babuska,fem-gockenbach};
\item {\it Neural networks (NN)} applied to solving ordinary and partial
differential
equations~\citep{1998-LagarisLikasFotiadis,Sirignano-Spiliopoulos-2018}.%
	\footnote{
	Two chapters on the applications of neural networks to solving
	differential equations were added after the textbook was published.
	These extra chapters are available for downloading as PDF files
	from the textbook's website~\citep{RosBookSite}.}
\end{enumerate}

In Spring 2021, based on the students' backgrounds and interests,
we selected the topics~1, 4, and~5 from the above list.   In view
of the interdependencies of the contents of the various chapters
as seen in Figure~\ref{fig:dependencies}, we included several prerequisite
chapters, which led to the following complete course syllabus:

\begin{figure}
	\centering
	\hspace*{-2cm}%
	\begin{tikzpicture}[x=0.8cm,y=-0.8cm, xscale=0.8, yscale=0.6,
			every path/.style={draw=lightgray, very thin}]

		\newcommand{\deptabr}{23}	
		\newcommand{\deptabc}{17}	

		\foreach \r in {2,...,\deptabr} {
			\ifnum \r > \deptabc
				\fill[black!7] (0,\r) rectangle +(\deptabc,-1);
			\else
				\fill[black!7] (0,\r) rectangle +(\r-1,-1);
			\fi
		}

		\foreach \c in {0,...,\deptabc}
			\draw (\c,0) -- +(0,\deptabr);

		\foreach \r in {2,...,\deptabr} {
			\ifnum \r > \deptabc
				\draw (0,\r) -- +(\deptabc,0);
			\else
				\draw (0,\r) -- +(\r,0);
			\fi
		}

		\draw (0,0) -- (\deptabc,0);

		
		\path[every node/.style={left=0.4em, text width=4.0cm-0.4em, inner sep=0pt, align=right}]
			(0,0.5) node (H-xmalloc)		{\bf 7. \sl Xmalloc}
			++(0,1) node (H-array-h)		{\bf 8. \sl array.h}
			++(0,1) node (H-fetchline)		{\bf 9. \sl Fetch line}
			++(0,1) node (H-random)			{\bf 10. \sl Random}
			++(0,1) node (H-sparse)			{\bf 11. \sl Sparse matrices}
			++(0,1) node (H-umfpack)		{\bf 12. \sl Umfpack}
			++(0,1) node (H-wavelets)		{\bf 13. \sl Wavelets}
			++(0,1) node (H-image-io)		{\bf 14. \sl Image I/O}
			++(0,1) node (H-image-anal)		{\bf 15. \sl Image analysis}
			++(0,1) node (H-linked-lists)	{\bf 16. \sl Linked lists}
			++(0,1) node (H-evolution)		{\bf 17. \sl Evolution}
			++(0,1) node (H-nelder-mead)	{\bf 18. \sl Nelder--Mead}
			++(0,1) node (H-trusses)		{\bf 19. \sl Trusses}
			++(0,1) node (H-fd1)			{\bf 20. \sl FD in 1D}
			++(0,1) node (H-pme)			{\bf 21. \sl Porous medium}
			++(0,1) node (H-gauss-quad)		{\bf 22. \sl Gauss quadrature}
			++(0,1) node (H-meshing)		{\bf 23. \sl Triangulation}
			++(0,1) node (H-twb-quad)		{\bf 24. \sl TWB quadrature}
			++(0,1) node (H-fem1)			{\bf 25. \sl FEM 1}
			++(0,1) node (H-fem2)			{\bf 26. \sl FEM 2}
			++(0,1) node (H-nn-odes)		{\bf 27. \sl NN -- ODEs}
			++(0,1) node (H-nn-pdes)		{\bf 28. \sl NN -- PDEs}
			++(0,1) node (H-barycentric)	{\bf A.  \sl Barycentric}
		;

		\path[every node/.style={right, text width=4.0cm, align=left, rotate=65}]
		 (0.5,-0.3) node (V-xmalloc)		{\bf 7. \sl Xmalloc}
			++(1,0) node (V-array-h)		{\bf 8. \sl array.h}
			++(1,0) node (V-fetchline)		{\bf 9. \sl Fetch line}
			++(1,0) node (V-random)			{\bf 10. \sl Random}
			++(1,0) node (V-sparse)			{\bf 11. \sl Sparse matrices}
			++(1,0) node (V-umfpack)		{\bf 12. \sl Umfpack}
			++(1,0) node (V-wavelets)		{\bf 13. \sl Wavelets}
			++(1,0) node (V-image-io)		{\bf 14. \sl Image I/O}
			++(1,0) node (V-linked-lists)	{\bf 16. \sl Linked lists}
			++(1,0) node (V-nelder-mead)	{\bf 18. \sl Nelder--Mead}
			++(1,0) node (V-fd1)			{\bf 20. \sl FD in 1D}
			++(1,0) node (V-gauss-quad)		{\bf 22. \sl Gauss quadrature}
			++(1,0) node (V-meshing)		{\bf 23. \sl Triangulation}
			++(1,0) node (V-twb-quad)		{\bf 24. \sl TWB quadrature}
			++(1,0) node (V-fem1)			{\bf 25. \sl FEM 1}
			++(1,0) node (V-nn-odes)		{\bf 27. \sl NN -- ODEs}
			++(1,0) node (V-barycentric)	{\bf A.  \sl Barycentric}
		;

		\LARGE
		\path[red!70!black]
			(H-array-h -| V-xmalloc.west)			node {$\bullet$}

			(H-random -| V-xmalloc.west)			node {$\bullet$}
			(H-random -| V-array-h.west)			node {$\bullet$}

			(H-sparse -| V-xmalloc.west)			node {$\bullet$}
			(H-sparse -| V-array-h.west)			node {$\bullet$}

			(H-umfpack -| V-sparse.west)			node {$\bullet$}
			(H-umfpack -| V-xmalloc.west)			node {$\bullet$}
			(H-umfpack -| V-array-h.west)			node {$\bullet$}

			(H-linked-lists -| V-xmalloc.west)		node {$\bullet$}

			(H-wavelets -| V-xmalloc.west)			node {$\bullet$}
			(H-wavelets -| V-array-h.west)			node {$\bullet$}

			(H-image-io -| V-xmalloc.west)			node {$\bullet$}
			(H-image-io -| V-array-h.west)			node {$\bullet$}

			(H-image-anal -| V-xmalloc.west)		node {$\bullet$}
			(H-image-anal -| V-array-h.west)		node {$\bullet$}
			(H-image-anal -| V-image-io.west)		node {$\bullet$}
			(H-image-anal -| V-wavelets.west)		node {$\bullet$}

			(H-evolution -| V-fetchline.west)		node {$\bullet$}
			(H-evolution -| V-xmalloc.west)			node {$\bullet$}
			(H-evolution -| V-array-h.west)			node {$\bullet$}
			(H-evolution -| V-random.west)			node {$\bullet$}
			(H-evolution -| V-linked-lists.west)	node {$\bullet$}

			(H-nelder-mead -| V-xmalloc.west)		node {$\bullet$}
			(H-nelder-mead -| V-array-h.west)		node {$\bullet$}

			(H-trusses -| V-fetchline.west)			node {$\bullet$}
			(H-trusses -| V-xmalloc.west)			node {$\bullet$}
			(H-trusses -| V-array-h.west)			node {$\bullet$}
			(H-trusses -| V-linked-lists.west)		node {$\bullet$}
			(H-trusses -| V-nelder-mead.west)		node {$\bullet$}

			(H-fd1 -| V-xmalloc.west)				node {$\bullet$}
			(H-fd1 -| V-array-h.west)				node {$\bullet$}

			(H-pme -| V-xmalloc.west)				node {$\bullet$}
			(H-pme -| V-array-h.west)				node {$\bullet$}
			(H-pme -| V-fd1.west)					node {$\bullet$}

			(H-meshing -| V-xmalloc.west)			node {$\bullet$}
			(H-meshing -| V-array-h.west)			node {$\bullet$}

			(H-twb-quad -| V-xmalloc.west)			node {$\bullet$}
			(H-twb-quad -| V-array-h.west)			node {$\bullet$}
			(H-twb-quad -| V-barycentric.west)		node {$\bullet$}
			(H-twb-quad -| V-meshing.west)			node {$\bullet$}

			(H-fem1 -| V-xmalloc.west)				node {$\bullet$}
			(H-fem1 -| V-array-h.west)				node {$\bullet$}
			(H-fem1 -| V-sparse.west)				node {$\bullet$}
			(H-fem1 -| V-umfpack.west)				node {$\bullet$}
			(H-fem1 -| V-barycentric.west)			node {$\bullet$}
			(H-fem1 -| V-meshing.west)				node {$\bullet$}
			(H-fem1 -| V-twb-quad.west)				node {$\bullet$}

			(H-fem2 -| V-xmalloc.west)				node {$\bullet$}
			(H-fem2 -| V-array-h.west)				node {$\bullet$}
			(H-fem2 -| V-sparse.west)				node {$\bullet$}
			(H-fem2 -| V-umfpack.west)				node {$\bullet$}
			(H-fem2 -| V-barycentric.west)			node {$\bullet$}
			(H-fem2 -| V-meshing.west)				node {$\bullet$}
			(H-fem2 -| V-gauss-quad.west)			node {$\bullet$}
			(H-fem2 -| V-twb-quad.west)				node {$\bullet$}
			(H-fem2 -| V-fem1.west)					node {$\bullet$}

			(H-nn-odes -| V-xmalloc.west)			node {$\bullet$}
			(H-nn-odes -| V-array-h.west)			node {$\bullet$}
			(H-nn-odes -| V-nelder-mead.west)		node {$\bullet$}

			(H-nn-pdes -| V-xmalloc.west)			node {$\bullet$}
			(H-nn-pdes -| V-array-h.west)			node {$\bullet$}
			(H-nn-pdes -| V-nelder-mead.west)		node {$\bullet$}
			(H-nn-pdes -| V-nn-odes.west)			node {$\bullet$}

		;
	\end{tikzpicture}
	\caption{
		To find the prerequisites of a chapter, find the chapter title
		along the left edge, go across horizontally to the bullet marks, and then go
		vertically to the prerequisite chapters.  For instance,
		{\bf Chapter~23: \sl Triangulation}
		depends on
		{\bf Chapter~7: \sl Xmalloc}, 
		and
		{\bf Chapter~8:  \sl array.h}.
		Chapters prior to Chapter~7 are not listed; these
		provide a general background for the entire book and should be
		considered prerequisites for everything.
	}
	\label{fig:dependencies}
\end{figure}
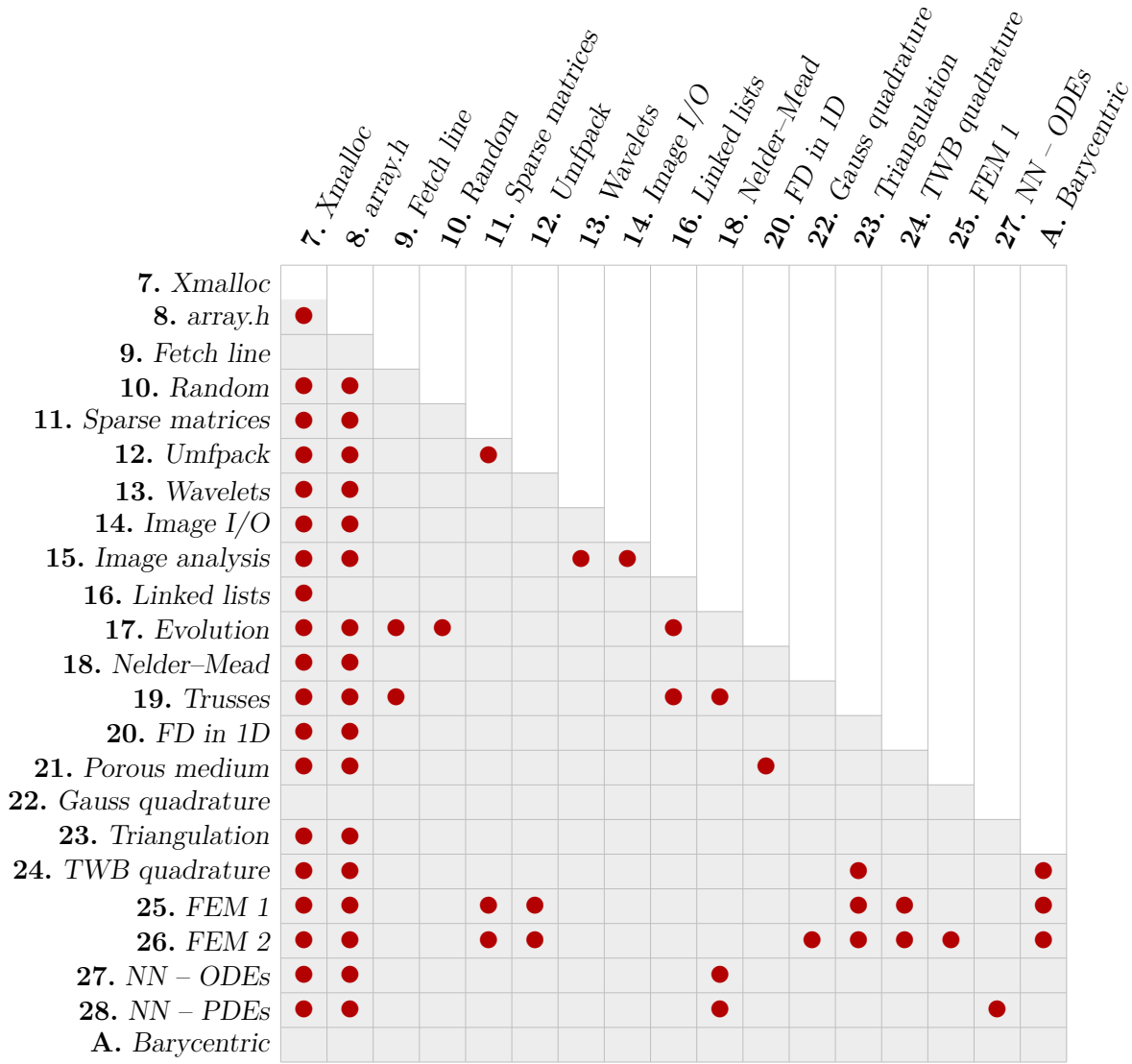

\begin{description}

\item[\bf Chapter 7: Memory allocation.]\mbox{}\\
	Dynamic allocation of memory lies at the core of practically
	every scientific computing program in C\@. This chapter
	sets up a safe ``wrapper'' around the \emph{C Standard Library's}
	\verb|malloc()| function to verify the success of the calls 
	to \verb|malloc()| and to take appropriate action in case of failure.

\item[\bf Chapter 8: Vectors and matrices.]
	Almost all of the book's projects call for allocating and freeing
	of memory for vectors and matrices of dynamically determined dimensions.
	This chapter puts the C preprocessor to good use by producing generic
	macros that perform the arduous task of constructing and freeing vectors
	and matrices of arbitrary dimensions.

\item[\bf Chapter 18: The Nelder--Mead simplex method.]
	The Nelder--Mead algorithm is a robust method for locating the
	minima of functions of the type $f : R^n \to R$.  The great
	appeal of the Nelder--Mead algorithm lies in the fact that it
	requires neither the differentiability of $f$, nor a knowledge of
	its gradient.  That should be contrasted against the 
	conjugate gradient algorithm and its variants, which do call for
	the knowledge of the often mathematically and computationally expensive
	calculation of the gradient.

\item[\bf Chapter 20: Finite difference methods.]
	This chapter introduces several finite difference algorithms
	for solving the one dimensional time dependence heat equation
	$\frac{\partial u}{\partial t} = \frac{\partial^2 u}{\partial x^2}$.
	The methods include the well-known explicit and implicit Euler
	and the Crank--Nicolson algorithms, as well as the lesser known
	but very versatile \emph{Seidman scheme}.

\item[\bf Chapter 11: Sparse matrices.]
	As a first step toward the implementation of the finite element methods
	of chapters~25 and~26, this chapter introduces the concept of
	\emph{Compressed Column Storage} (CCS) form for memory-efficient
	storage of large sparse
	matrices. The ideas discussed here form the foundation for understanding
	and appreciating the details of the next chapter's \UMFPACK\ library.

\item[\bf Chapter 12: The UMFPACK library.]
	The \UMFPACK\ is an open source library that incorporates state-of-the-art algorithms
	for solving algebraic linear systems of equations $Ax=b$, where
	$A$ is an $n\times n$ matrix and $b$ is an $n$-vector.  \UMFPACK\ is particularly efficient
	when solving large sparse linear systems. That is exactly the type of system
	that arises when solving partial differential equations through
	the finite element method.

\item[\bf Chapter 23: Triangulation of polygons.]
	\ \emph{Meshing} or \ \emph{triangulation}
	is a critical step in solving partial differential
	equations by a finite element method on polygonal domains. It
	involves partitioning of the domain into a union of non-overlapping
	triangles.  It can be shown that the finite elements discretization
	error is the smallest when the triangulation involves no triangles
	with sharp angles. The award-winning and open-source
	\Triangle\ library produces
	quality triangulations that come close to the best that can be achieved
	within the limitations of the domain's constraints.

\item[\bf Appendix A: Barycentric coordinates.]
	The geometry of the individual triangles in a domain's triangulation
	is best expressed in barycentric coordinates.  This appendix
	provides a quick overview of the barycentric coordinates.
	The material here is essential for understanding the developments
	in chapters~24 through~26.

\item[\bf Chapter 24: Integration on triangles.]
	The entries of the matrix $A$ and vector $b$
	noted in the earlier paragraph on \UMFPACK\
	are produced through integration over individual triangles of a
	triangulated domain. This chapter presents a relatively recent
	accurate and efficient method for numerical integration on
	triangles~\citep{2007a-taylor-etal,2007b-taylor-etal}.

\item[\bf Chapter 25: Finite elements (part~1).]
	This chapter introduces a basic finite element scheme with linear
	basis elements for solving the Poisson problem
	$
		\frac{\partial^2 u}{\partial x^2}
		+
		\frac{\partial^2 u}{\partial y^2}
		+ f(x,y) = 0,
	$
	on an arbitrary polygonal domain, possibly with polygonal holes,
	and null Dirichlet
	data.  The solution involves triangulating the domain with Chapter~23's \Triangle\
	library, calculating the stiffness matrix by integrating over the
	triangles through Chapter~24's algorithms, and solving the resulting linear system with Chapter~12's \UMFPACK.

\item[\bf Chapter 22: Gaussian quadrature.]
	This chapter develops tools for integrating functions of a single
	variable through Gaussian quadrature. The material here can be
	useful on its own---see~\citep{atkinson,kincaid-cheney} for in-depth
	coverage---but our main objective is directed toward
	Chapter~26 where we will be solving partial differential equations with prescribed boundary fluxes.

\item[\bf Chapter 26: Finite elements (part~2).]
	This chapter extends Chapter~25's elementary treatment of
	finite elements to the elliptic equation
	$$\nabla \cdot ( \eta(x,y) \nabla u) + f(x,y) = 0,$$
	where $\nabla$ is the gradient operator and $\eta(x,y)$ is
	a prescribed, and generally variable, diffusion coefficient.
	Arbitrary Dirichlet and Neumann data may be specified on
	the boundary.  The finite elements formulation then calls
	for the integration of the Neumann data on the domain's boundary.
	The integration is performed through Gaussian quadrature
	introduced in Chapter~22.
\end{description}

\begin{figure}
	\centering
		\includegraphics[width=0.33\textwidth,angle=90]{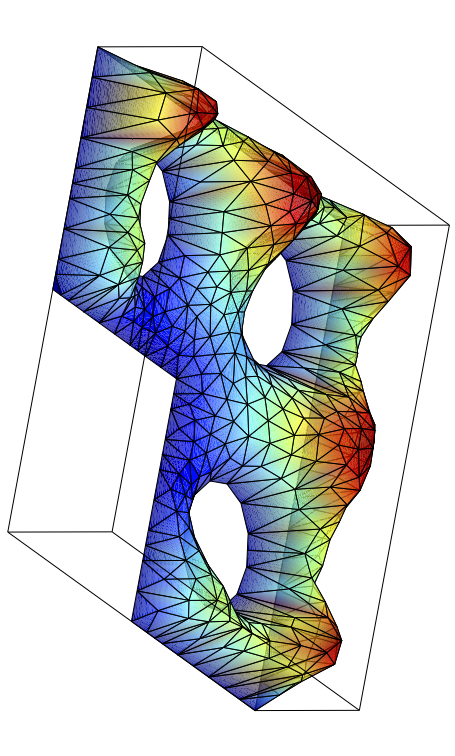}
		\qquad
		\includegraphics[width=0.25\textwidth]{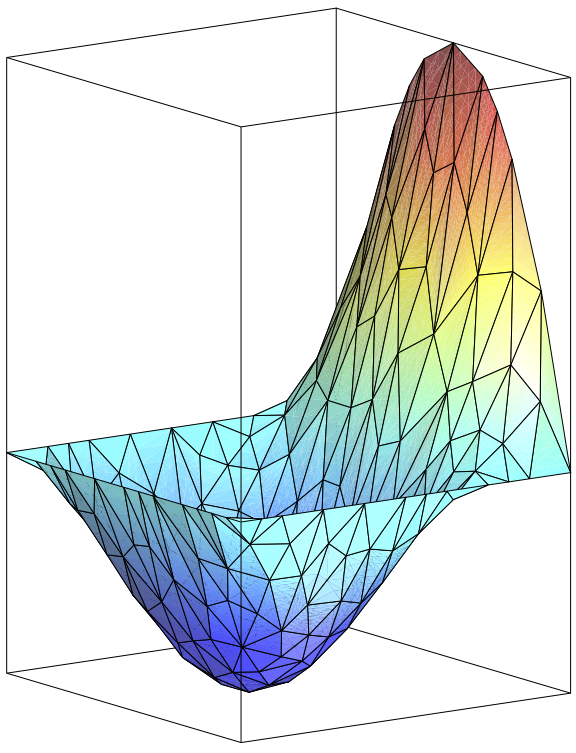}
	\caption{Samples of solutions of Poisson's partial differential
	equation produced by the finite elements solver.
	On the left we have an L-shaped domain with holes, and zero Dirichlet
	boundary data. On the right we have a square domain with zero
	Dirichlet data specified on three edges, and a prescribed flux on the
	fourth edge..}
	\label{fig:sample-fem-solutions}
\end{figure}

Chapter 26 is only briefly introduced at the end of the semester. Students are encouraged to 
take MATH~4373/5373: \textsl{Finite Element Methods (FEM)} in scientific computing, a new course developed 
for \UofA\ students. This course is under continuous development to provide a broad 
range of support to the students in science and engineering departments. Bangerth~\citep{ZarBan04} 
describes a practitioner's approach to using the principles of reflective writing and journaling to connect the 
material of the video lectures to student projects in a project-based course on FEM, MATH~676, offered 
at Texas A\&M University.  
\subsection{Performance Assessments}
\label{sect:perfAsses}

Performance assessments are based on 60\% assignments
and 40\% final project.  Assignments are designed to focus
on solving specific problems through the projects listed in 
section~\ref{sec:proj}. Each project's mathematical background
and the complex aspects of the required programming are explained in class.
The students are asked to
complete the missing parts of the project.
To receive full credit, the resulting programs
need to compile without warnings and errors, and run
successfully.

The final projects are designed to utilize visual, verbal,
and written presentation skills---all essential ingredients for
practical and successful research.  For students who are involved in research outside
of this course, effort is made to pick projects that utilize
algorithms that may be applicable to their research.
For others, specific projects are offered and assigned by
the instructor.  

The objectives of this course are achieved by completing the final projects. 
Considering the students' backgrounds, the final projects are chosen from 
a wide range of applications from variable stars to self-driving cars.  
The student with a mathematics major (see Figure~\ref{fig:studentsBack}) chose his final project 
on the graph isomorphism problem which is a computational problem in graph theory
consisting of understanding and implementing an efficient algorithm to determine whether two graphs are isomorphic. 
The student with a double major in Computer Science and Computer Engineering and Mathematics completed a final
project on reinforcement learning for self-driving cars which rely on sensor inputs to perceive the environment 
and move safely without human interaction. 

The instructor meets the students twice to assign a project related to their interests and 
follow-up on progress made before the final presentations. 
At the beginning of the semester, each student gives a short presentation to outline his/her research interest. 
At the middle of the semester, the student gives a short presentation to show his/her progress on the project.
The students write a two page final project report using the sample LaTeX template provided by the instructor.  
This report includes sections on the problem description, the algorithm to solve the problem, and 
the implementation details of the algorithm outlining the results that they have achieved.  
In additional to this written report, the students give a 15--20 minute presentations at the end of the semester. 

Each student's opinion matters, therefore the students are asked to fill out and submit 
the peer evaluation and feedback form that includes 10 questions.  Students judge and evaluate the presentations of each peer
based on the presenter's explanation of the topic. The presentation is rated on a five point scale range 
from poor~=~1,  fair~=~2, good~=~3,  very good~=~4, and excellent~=~5. 
The poor-to-excellent rating scale provides a measure of the student's performance. 
\begin{enumerate}
\item The presenter delivered the material in a clear and structured manner. 
\item The presenter was knowledgeable about the topic.  
\item The presenter maintained my interest during the presentation. 
\item The presenter answered questions effectively.   
\item The presenter was enthusiastic about the topic.   
\item The presenter was well organized and prepared. 
\item The presentation was concise and informative.    
\item The presentation contained practical examples and useful techniques that applied to current work.
\item The visual aids were effective.     
\item Overall, I would rate this presentation as.
\end{enumerate}
The instructor collects the evaluation forms and reviews them to assign the final grade
according to the percentage scale $A\geq 90, B \geq 80, C \geq 70, D \geq 60, F < 60$.

\subsection{Educational activities}
\label{sect:activities}

For the benefit of the graduate and undergraduate students,
the SIAM\footnote{\sl Society for Industrial and Applied
Mathematics} student chapter at the \UofA\ holds seminars and lecture series to provide a learning
environment outside of the existing classes specifically to develop
and improve the students' mathematical and programming skills.
The SIAM seminars help students to connect with faculty and l
earn more about the ongoing projects in various sciences and 
engineering departments. 
The lecture series are designed to provide extra help to students 
who have had no or very little experience in programming 
in C and using high performance computing systems.

We begin the semester with a lecture series that introduces
programming in C in the Linux environment. 
We explain fundamental UNIX/Linux commands and how to use text
editors such as \texttt{vim},  \texttt{nano},  and \texttt{pico}.
Working on a command-line in a terminal emulator, and reading and editing
programs in a text editor is a new experience for most of the students.
With theory and hands-on exercises as a part of the computational activities,
the students learn the basics in three 50-minute lectures. The lecture series
are used to complement the weekly course activities and serve as an infrastructure
for the course.

A vital pedagogical need is the monitoring of students to verify that 
they follow every step of the process to fully comprehend the material. 
We have hands-on sessions after the lectures. Students work in groups 
on the computer-based solution of mathematical problems that range 
from simple to complex. Whenever possible, we pair an undergraduate 
student with a graduate student to provide mentorship and facilitate solving the problems. 

\section{Undergraduate Opportunities}
\label{sect:internships}

In this section, we present the undergraduate research opportunities made available annually in the 
\textsl{Computational and Applied Mathematics} (CAM) research group at
the \UofA, to help and encourage
undergraduate students to engage in computational research.
The positions are created under the auspices of \textsl{Shodor} (\url{http://www.shodor.org/}),,
a non-profit foundation and a national resource
for computational science education, with a mission
to improve mathematics and science education through
the effective use of modeling and simulation technologies---``computational science''.

Under Shodor, the National Computational Science Institute (NCSI)(\url{http://computationalscience.org}) 
provides workshops on computational science for educators at all levels to give
them ideas and resources to use in their classrooms.
The NCSI is responsible running undergraduate student
programs such as the \textsl{Blue Waters internship} 
(\url{https://bluewaters.ncsa.illinois.edu/internships})  and 
XSEDE--EMPOWER 
(Extreme Science and Engineering Discovery Environment--Expert
Mentoring Producing Opportunities for Work, Education, and Research)
to increase the number of students interested in developing 
computational skills 
(\url{http://computationalscience.org/xsede-empower}). 

A faculty member at an U.S.  academic institution who would like to mentor an
undergraduate student submits a research proposal to the NCSI program to 
create a position in his/her research group.
The program provides a stipend for students per quarter, semester, summer or longer,
depending on the expected level of effort.
Undergraduate students from any 
U.S. degree-granting institutions are matched
with a mentor who has a project that contributes to the work of XSEDE~\citep{XSEDE14}. 
There are three tiers of participation for students, depending on their skill levels: 
\emph{learner, apprentice, and intern} (\url{http://computationalscience.org/xsede-empower}).
The \emph{learner level} is for a student who has no experience in scientific computing. 
The student spends time developing necessary skills to contribute 
to the work through online tutorials, workshops, and self learning in programming. 
At the \emph{apprentice level},  the student begins to transform the knowledge into skills, 
and has the opportunity to apply the new skills with some additional training 
in debugging and performance tools to perform the assigned tasks. 
After completing these two levels in two semesters, the students 
are accepted as \emph{interns} in the program. 

The CAM group leader has created several projects 
for the undergraduate students 
who wish to gain research experience during their studies. 
The goal of these projects is to engage the students in petascale/exascale 
computing research in the areas of modeling and simulations; 
numerical methods and performance optimization. 
As part of the Blue Waters (\url{https://bluewaters.ncsa.illinois.edu/}) efforts to motivate and 
train the next generation of supercomputing researchers, 
two \UofA\ undergraduate students,  
Edwards (major in Mathematics, minor in Physics/Computer Science) and 
McGarigal (major in Mechanical Engineering, minor in Mathematics) 
were elected as the 2018--2019 Blue Waters Student Interns.
The mentor of two undergraduate students in this year-long internship 
was responsible for teaching and introducing the use of HPC 
for the numerical simulations of flow problems in the area of computational fluid dynamics. 
To be able to work on the proposed project, 
the students had to develop skills in programming languages;  
parallel programming models (MPI on distributed memory, OpenMP on shared memory); 
scientific visualization (VisIt); 
performance analysis tools (TAU, CPMAT); and usage of the Blue Waters systems environment.
The usage of HPC systems requires familiarity with compilers on Linux systems, 
submitting batch scripts, and running and debugging programs. 
The visualization of simulation results was done through
the open source, interactive, scalable, tool \textsl{VisIt} ({\url{https://visit.llnl.gov/}). 
The performance analysis of the software written during the internship 
was done using several different profiling tools, such as
GNU profiling tool (gprof), the Cray Performance Measurement and Analysis Tools (CPMAT),
and Tuning and Utility Analysis (TAU). 
With the help of these profiling tools, the students were able to identify 
performance bottlenecks in the application code, visualize the 
data, and achieve performance improvements through hybrid (MPI+OpenMP) programming. 
This modified hyper-threading version of the code was set up in a way that multiple 
message passing interface (MPI) processes handle the interface propagation, whereas multiple 
OpenMP threads handle the higher order weighted essentially non-oscillatory numerical scheme. 
This undergraduate research project was selected for publication in 
the Journal of Computational Science Education~\citep{KamEdwMcG21}.

The directed reading course--MATH~400V--, designed to help the students who have no 
or very little experience in programming, is used as the building blocks 
for the two-week crash course of MATH~4343 on the C programming language,
and the programming environment on the university's high performance
computing system. Students who complete MATH~4343 with grades of ``A''
are accepted to the CAM group to perform more independent work and 
to become more fully engaged in research. 

After the research experience in the CAM group, 
almost all students choose to pursue graduate studies and 
continue to work on CSE research projects. 
For instance; Edwards (Blue Waters Intern) was one of ten students 
accepted to the Oak Ridge National Laboratory's \textsl{Pathways to Computing 
Internship Program} to learn and develop the next-generation explicit 
methods for radiation transport in astrophysics and 
explore programming models for GPU supported on the fastest supercomputer 
in the world, Summit~\citep{top500}. In Fall 2021, she started her graduate studies 
in the Computational and Mathematical Sciences at the Florida State University.  
Drinh (XSEDE 2019 learner) was accepted to the University of Missouri-Kansas City.   
De-La Cruz (XSEDE 2021 apprentice and intern) intends to pursue a Ph.D. in CSE
after graduating from University of Ozarks in 2021.

\section{Evaluation of the Course}
\label{sect:eval}

The course is evaluated based on the six questions listed in
Table~\ref{courseEval}.  It shows the percentage of students 
who responded to the questions with SA, A, U, D, SD, which stand 
for strongly agree, agree,  undecided, disagree and strongly disagree respectively.
The instructor performance is evaluated based on the three questions
listed in Table~\ref{instrEval}. 
The students rate this course and the instructor as excellent (50\%) and good (50\%). 
See Table~\ref{rateCourseInstructor}.

\begin{table}[!h]
\tbl{Course evaluation survey. Responses: [SA]~Strongly Agree=5, [A]~Agree=4, [U]~Undecided=3,
 [D]~Disagree=2, [SD]~Strongly Disagree=1. }
{\begin{tabular}{ |p{7cm}|ccccc| }
\hline
 \multirow{2}{*}{Questions} & \multicolumn{5}{c|}{Responses(\%)} \\ \cline{2-6}
    & SA & A & U & D & SD \\ \hline
 Q1.  Assignments are related to goals of this course.  & 25\% & 75\% & 0 & 0 & 0 \\ \hline
 Q2. The teaching methods used in this course enable me to learn.  & 50\% & 50\% & 0 & 0 & 0 \\ \hline
 Q3. The stated goals of this course are consistently pursued. & 50\% & 50\% & 0 & 0 & 0 \\ \hline
 Q4. I actively participate in class activities and discussions. & 25\% & 50\% & 25\% & 0 & 0 \\ \hline
 Q5.  I put much effort into this course.  & 50\% & 25\% &  25\% & 0 & 0 \\ \hline
 Q6. My problem-solving abilities improved because of this course.  & 25\% & 75\% & 0 & 0 & 0 \\ \hline
\end{tabular}}
\label{courseEval}
\end{table}

\begin{table}[!h]
\tbl{Instructor evaluation survey. Responses: [SA]~Strongly Agree=5, [A]~Agree=4, [U]~Undecided=3,
 [D]~Disagree=2, [SD]~Strongly Disagree=1. }
{\begin{tabular}{ |p{7cm}|ccccc| }
\hline
 \multirow{2}{*}{Questions} & \multicolumn{5}{c|}{Responses(\%)} \\ \cline{2-6}
    & SA & A & U & D & SD \\ \hline
 Q1. My instructor displays a clear understanding of the topic.  & 75\% & 25\% & 0 & 0 & 0 \\ \hline
 Q2. My instructor is readily available for consultation.  & 75\% & 25\% & 0 & 0 & 0 \\ \hline
 Q3. My instructor explains difficult material clearly.  & 50\% & 50\% & 0 & 0 & 0 \\ \hline
\end{tabular}}
\label{instrEval}
\end{table}

\begin{table}[!h]
\tbl{Rate the course and the instructor. Responses: [E]~Excellent=5, [G]~Good=4, [F]~Fair=3, [P]~Poor=2, [VP]~Very Poor=1 . }
{\begin{tabular}{ |l|ccccc| }
\hline
 \multirow{2}{*}{Questions} & \multicolumn{5}{c|}{Responses(\%)} \\ \cline{2-6}
											& E & G & F & P & VP \\ \hline
 Overall, I would rate this course as: & 50\% & 50\% & 0 & 0 & 0 \\ \hline
 Overall, I would rate this instructor as:& 50\% & 50\% & 0 & 0 & 0 \\ \hline
\end{tabular}}
\label{rateCourseInstructor}
\end{table}

\section{Concluding Remarks and Future Plans}
\label{sect:conclusion}

In this paper, we have introduced the MATH~4343 ``Introduction to Scientific Computing", 
a course designed and first taught in Spring 2021 at the \UofA, 
to engage undergraduate students in research and to train them 
as future computational scientists and engineers. 
The focus of this interdisciplinary undergraduate course is
on formulating a diverse set of challenging interdisciplinary problems into
mathematical algorithms, and implementing these algorithms 
on high performance computing systems. 
This helps develop the students' skill in mathematics, programming, 
computational modeling, simulation, visualization, and eventually
prepares them for graduate studies and 
future careers in research and industry.  

In Spring 2022, the number of students enrolled increased by 70\% relative to
that of the Spring 2021 semester. In addition, 
four graduate students (one geosciences and three mathematics) have sought
permission to enroll in this undergraduate course.
We are considering offering the equivalent of this course
in the future at the graduate level,
and begin to target graduate students 
not only in applied mathematics but also in other science and engineering programs.

We envision that the contents of the course will evolve
in response to the shifting interests of the students,
as well as those of the instructors.  In view of the recent
developments on the applications of neural network models
to machine learning and ``big data'', we intend to introduce
projects on neural networks in MATH~4343 course in the immediate future.
As it was briefly noted in section~\ref{sec:proj}, two new
chapters on neural networks have been added to the course's textbook,
and therefore we have ready-made material to work with.
The two chapters highlight the application of neural networks
to solving boundary value problems for nonlinear ordinary
and partial differential equations.  The mesh-free approach
developed there enables students to solve significantly complex
problems which minimal effort.  Here, for instance, is a boundary
value problem for a nonlinear PDE on a non-trivial two-dimensional domain:
\begin{equation} \label{eq:nn-bvp}
\begin{aligned}x
	\frac{\partial^2 u}{\partial x^2}
	+ 
	\frac{\partial^2 u}{\partial y^2}
	+ \frac{4}{1+u^2} &= 0	&&\quad\mbox{in } \Omega,
	\\
	u &= 0 &&\quad\mbox{on } \partial\Omega.
\end{aligned}
\end{equation}
Figure~\ref{fig:nn-solution} shows the domain $\Omega$ and
the graph of the solution $u(x,y)$ produced by our neural network.
Implementing the neural network can be done fairly quickly since,
as can be seen in Figure~\ref{fig:dependencies},
there are only four prerequisite chapters to apply neural networks 
for solving ordinary and partial differential equations.
The construction of the domain $\Omega$ and the enforcing of the boundary
conditions, is made possible through Rvachev's R-functions~\citep{1995-Rvachev-Sheiko,2007-Shapiro} 
which are introduced in the course's textbook's Chapter~28.

\begin{figure}
	\centering
	\mbox{
	\includegraphics[width=0.25\textwidth]{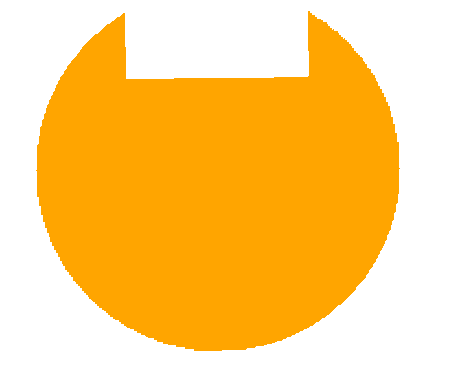}
	\qquad
	\includegraphics[width=0.45\textwidth]{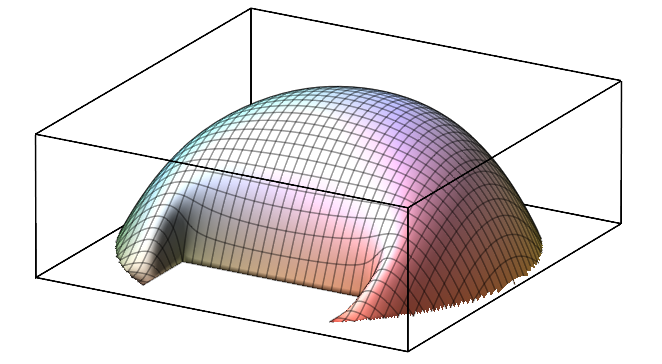}
	}
	\caption{The domain $\Omega$ (on the left) and the graph of solution
	(on the right) of the nonlinear boundary value problem~\eqref{eq:nn-bvp}.
	The grid on the surface is drawn for visualization purposes only; the
	neural network solver's algorithm is mesh-free.
	}
	\label{fig:nn-solution}
\end{figure}

\section*{Acknowledgements}
\label{sect:acks}
This research is supported by the \textsl{Arkansas High Performance Computing Center} 
which is funded through multiple National Science Foundation grants and the 
Arkansas Economic Development Commission. 
Use of computational facilities on the Pinnacle at the University of Arkansas is gratefully acknowledged.  
This work was supported by a grant from the Shodor Education Foundation through Blue Waters Student Internship Program. This research is part of the Blue Waters sustained-petascale computing project, which is supported by the National Science Foundation (awards OCI-0725070 and ACI-1238993) and the state of Illinois. Blue Waters is a joint effort of the University of Illinois at Urbana-Champaign and its National Center for Supercomputing Applications.
This work used the Extreme Science and Engineering Discovery 
Environment (XSEDE) bridges.psc.xsede.org at the 
Pittsburgh Supercomputing Center through allocation DMS190029.

\end{document}